\documentclass{amsart}
\date{March 17, 2003}
\usepackage{amsmath,amsfonts}

\newtheorem{lemma}{Lemma}

\newtheorem{theorem}{Theorem}
\newtheorem{corollary}{Corollary}

\newcommand{\bk}{\mathbf{k}}
\newcommand{\bp}{\mathbf{p}}
\newcommand{\bq}{\mathbf{q}}
\newcommand{\br}{\mathbf{r}}
\newcommand{\bx}{\mathbf{x}}
\newcommand{\by}{\mathbf{y}}

\newcommand{\gH}{\mathfrak{H}}
\newcommand{\gS}{\mathfrak{S}}

\newcommand{\rc}{\mathrm{c}}

\newcommand{\cz}{\mathbb{C}} 
\newcommand{\rz}{\mathbb{R}} 

\newcommand{\Hh}{\mathbb{H}}
\newcommand{\Qv}{{Q^\mathrm{\varphi}}}
\newcommand{\dK}{{\int_{-\infty}^{\infty}d\eta}}
\newcommand{\ide}{\frac 1{D^0 + i \eta}}
\newcommand{\idve}{\frac 1{\dv + i \eta}}
\newcommand{\W}{\mathbb{W}}

\newcommand{\tq}{{\tilde Q}}
\newcommand{\dv}{D^\varphi}
\newcommand{\la}{\langle}
\newcommand{\ra}{\rangle}
\newcommand{\Pvp}{P_+^\mathrm{\varphi}}
\newcommand{\Pvn}{P_-^\mathrm{\varphi}}
\newcommand{\Pop}{P_+^{0}}
\newcommand{\Pon}{P_-^{0}}
\newcommand{\Hp}{\gH_+^\varphi}
\newcommand{\Hn}{\gH_-^\varphi}
\newcommand{\alp}{\boldsymbol{\alpha}}

\def\tr{\mathop{\rm tr}\nolimits} 
\def\Tr{{\rm tr}_{\cz^4}}

\title[Vacuum Polarization]{Non-Perturbative Mass and Charge Renormalization in
  Relativistic No-Photon Quantum Electrodynamics}

\thanks{\copyright
    2003 The authors. Reproduction of this article for non-commercial
    purposes by any means is permitted.\\ C.H. has been supported by a
    Marie Curie Fellowship of the European Community program
    ``Improving Human Research Potential and the Socio-economic
    Knowledge Base'' under contract number HPMFCT-2000-00660.  Both
    authors acknowledge partial support through the European Union's IHP
    network Analysis \& Quantum HPRN-CT-2002-00277}

\author[C. Hainzl]{Christian Hainzl}
\address{Mathematik, Theresienstra\ss e 39, 80333 M\"unchen,
  Germany. Address as of March 1, 2003: CEREMADE, Universit\'e
  Paris-Dauphine, Place du Mar\'echal de Lattre de Tassigny, F-75775 Paris
  Cedex 16, France.} \email{hainzl@ceremade.dauphine.fr}

\author[H. Siedentop]{Heinz Siedentop}
 \address{Mathematik,
  Theresienstra\ss e 39, 80333 M\"unchen, Germany.}
  \email{h.s@lmu.de}

\begin{document}

 \markboth{C. Hainzl, H.
  Siedentop}{Vacuum Polarization}

\begin{abstract}
  Starting from a formal Hamiltonian as found in the physics
  literature -- omitting photons -- we define a renormalized
  Hamiltonian through charge and mass renormalization. We show that
  the restriction to the one-electron subspace is well-defined. Our
  construction is non-perturbative and does not use a cut-off.

  The Hamiltonian is relevant for the description of the Lamb shift in
  muonic atoms.
\end{abstract}

\maketitle

\section{Introduction\label{s1}}

According to Dirac's hole theory the vacuum consists of electrons
which occupy the negative energy states of the free Dirac operator
(Dirac sea). Dirac postulated that their charge is not measurable.
However, if one introduces an external electric field, e.g., the field
of a nucleus, these electrons should rearrange, occupying the negative
energy states of the Dirac operator with the external electric field.
Physically speaking, the nucleus polarizes the vaccuum. (This
rearrangement may be interpreted as the creation of virtual
electron-positron pairs when expressed in terms of the free Dirac
operator.) In other words, the vacuum is polarized.  Dirac
\cite{Dirac1934} indicates that these polarization effects result in a
logarithmically divergent charge density, which cannot be neglected.
As a solution, he suggested that a momentum cut-off must be
introduced, since he expected that the Dirac equation would fail for
energies higher than $137 mc^2$.  In \cite{Dirac1934D} he changed this
train of thought and suggested that the infinities occurring should be
absorbed by a procedure which is now called charge renormalization. A
similar step was independently undertaken by Furry and Oppenheimer in
\cite{FurryOppenheimer1934} who circumvented the hole theory by
introducing annihilation and creation operators.

Heisenberg \cite{Heisenberg1934} clarified Dirac's picture and
generalized his approach extracting the physically relevant terms by
subtraction of an unambiguous infinite constant, at least to first
order in $\alpha$. Serber \cite{Serber1935} and Uehling
\cite{Uehling1935} gave detailed calculations (in first order of
$\alpha$). Uehling demonstrated that the vacuum polarization alters
the Coulomb potential of a charged particle resulting in the electron
being slightly more bound in the $s$-states (angular momentum $0$) of
hydrogenic atoms.  Later Weisskopf \cite{Weisskopf1936} gave a
thorough discussion of the physics involved in charge
renormalization.

After the experiments of Lamb and Retherford \cite{LambRetherford1947}
in 1947, which gave a much higher discrepancy concerning the hyperfine
structure of hydrogen, in addition to a different sign, than Uehling's
calculation showed, and the first explanation by Bethe
\cite{Bethe1947}, the insight into quantum electrodynamics (QED)
changed and the interaction with the radiation field turned out to be
the dominating part in describing the splitting of the energy levels
of hydrogenic atoms beyond the Dirac equation. Similar to vacuum
polarization, which was now treated together with the radiative
corrections, the photon interaction caused fundamental problems such
as infinities, which were ``removed'' -- at least in first order of
$\alpha$ -- by mass renormalization by Tomonaga, Schwinger, and
Feynman.  Eventually, Dyson ``succeeded'' with the renormalization
program to every order in $\alpha$.  Since then, QED has proven to be
of extraordinary predictive power.  (We refer the reader interested in
more historical details to Schweber \cite{Schweber1994}.)

But despite the predictive power of quantum electrodynamics, the
description in terms of perturbation theory causes great uneasiness
among mathematicians; a mathematically consistent formulation of QED
is still unknown; in fact Dyson \cite{Dyson1952} indicated that the
perturbation theory is divergent. A self-adjoint Hamiltonian for
QED is not known.

In the present paper we address a particular kind of singularities
arising in QED, namely those stemming from the vacuum polarization.
As opposed to the prevalent physics literature we will not use any
Feynman diagram but will rather construct a Hamiltonian (in Coulomb
gauge). This we have in common with the above cited early works in the
field.  However, the fact that we start from a formal Hamiltonian and
renormalize it \textit{non-perturbatively} distinguishes us from those
authors.

Altough our approach is rigorous, the resulting renormalization is far
from being of academic interest only. In fact, the restriction
$D_\mathrm{ren}$ of the fully renormalized Hamiltonian $\mathbb{H}$ to
the one-particle electron sector accounts already for a precise
description of the low energy levels of $\mu$-mesonic atoms where the
vacuum polarization effect dominates the radiative corrections by far,
since the Bohr orbits traverse the support of the polarization
potential in this case.  (See, e.g., Peterman and Yamaguchi
\cite{PetermanYamaguchi1959}, Glauber et al. \cite{Glauberetal1960},
Milonni \cite{Milonni1994}, Weinberg \cite{Weinberg1996}, and Greiner
et al \cite{Greineretal1985}.)

\section{Model\label{s2}}

In relativistic QED the quantized electron-positron field $\Psi(x)$,
which is an operator valued spinor, is written formally as
\begin{equation}
  \Psi(x)= a(x) + b^*(x)
\end{equation}
where $a(x)$ annihilates an electron at $x$ and $b^*(x)$ creates a
positron at $x$. (We use the notation that $x=(\bx,\sigma) \in \Gamma =
\rz^3\times \{1,2,3,4\}$, where $\sigma$ is the spin index and $\int dx$
denotes integration over $\rz^3$ and a summation over $\sigma$.)  The
underlying Hilbert space is given by $\gH = L^2(\Gamma)$.

The definition of a one-electron, respectively one-positron, state
will correspond to the positive, respectively negative, energy
solutions of the Dirac operator
\begin{equation}
  \dv = \alp \cdot\frac1i\nabla + \beta - \alpha\varphi
\end{equation}
in which $\alp,\beta$ denote the $4\times 4$ Dirac matrices. The
constant $\alpha$ is a positive real number, the Sommerfeld fine
structure constant which is approximately $1/137$. (We have picked
units in which the electron mass is equal to one.)

We will not assume that the nucleus is a point particle; we rather
associate with it a density $n\in L^1(\rz)$ whose integral gives the
atomic $Z$ number of the atom under consideration. For technical
convenience we assume $n$ to be a spherically symmetric Schwartz
function whose Fourier transform has compact support.

We remark: it is an experimental fact that the nucleus is not a point
particle but an extended object. In fact the numerical calculations of
the Lamb shift depend on the size of the support, which actually
limits the accuracy of numerical value of the calculation of the Lamb
shift because of the experimental uncertainty of the radius of the
nucleus (Weinberg \cite{Weinberg1996}, p. 593).  (A point nucleus
leads also to mathematical difficulties, since the renormalized
potential is more singular than the Coulomb potential, i.e., it could
not be controlled by the kinetic energy (see Uehling
\cite{Uehling1935} and Subsection \ref{coulomb} of this paper).  

The electric potential of the nucleus is given as
\begin{equation}
  \varphi =  |\cdot|^{-1} * n.
\end{equation}
An application of the Young inequality shows that the nuclear
potential 
\begin{equation}
  \label{eq:17}
  \varphi\in L^{3+\epsilon}(\rz^3)\cap L^\infty(\rz^3)
\end{equation} 
under our assumption on the nuclear density $n$ for any positive
$\epsilon$.  Moreover, by Newton's theorem
\begin{equation}
  \label{eq:2}
  0\leq \varphi(\bx)\leq Z/|\bx|.
\end{equation}

For completeness we note the following fact whose proof is obvious:
\begin{lemma}
  \label{DiracV}
  Fix $\alpha Z\in[0,1)$ and assume $n$ to be a non-negative Schwartz
  function with $\int n =Z$ and $\varphi=|\cdot|^{-1}*n$.  Then,
  $D^\varphi$ is selfadjoint with domain $H^1(\Gamma)$, i.e., the same
  domain as the free Dirac operator $D_0$. Moreover, if $n$ is
  spherically symmetric, $D^\varphi$ has no negative discrete spectrum
  and its $k$-th eigenvalues is bounded from below by the $k$-th
  eigenvalue of the Coulomb Dirac operator $D^{Z/|\cdot|}$.
\end{lemma}

We remark that Lemma \ref{DiracV} implies that the lowest eigenvalue
is positive.

Now, we can specify the electron and positron state spaces $\Hp$ and
$\mathcal{C}\Hn$ respectively: the orthogonal projection on $\Hp$ and
$\mathcal{C}\Hn$ are defined as
$\Pvp:=\chi_{[0,\infty)}(D^\varphi_{\alpha,m})$ and $\Pvn:=1-\Pvp$.  The
(anti-unitary) charge conjugation operator is given on $\gH$ by $(\mathcal{C}
\psi)(\bx) = i\beta \alpha_2 \overline{\psi(\bx)}$. In momentum space it acts
as $(\hat{\mathcal{C}} \hat \psi)(\bp)=i\beta \alpha_2 \overline{\hat
  \psi(-\bp)}$. (Here and in the following we follow the notation of Thaller
\cite{Thaller1992}; see also \cite{HelfferSiedentop1998}.)

Note that the Hilbert space can be written as the orthogonal sum
\begin{equation}
\gH = \Hp \oplus \Hn.
\end{equation}
Correspondingly $a^*(f)$ creates an electron in the state $\Pvp f$, whereas
$b^*(g)$ creates a positron with wave function $\mathcal{C} \Pvn g$. Note that
the definition of the operator $a$ and $b$ depends explicitly on the choice of
the potential $\varphi$.

The Hamiltonian for the non-interacting electron-positron field is
given by
\begin{equation}
  \label{eq:3}
  \mathbb{D}^\varphi = \int dx :\Psi^*(x) \dv \Psi(x):
\end{equation}
where $:\ :$ denotes normal ordering, i.e., anti-commuting of all
creation operators to the left of all annihilations operators ignoring
the anti-commutators.

Note, that for our renormalization procedure the choice of the
electron-positron subspaces as the positive and negative eigenspaces
of $\dv$ is crucial, in fact it is a choice already proposed by Dirac
\cite{Dirac1934}.

The creation and annihilation operators fulfill the canonical
anti-commutation relations
\begin{equation}
  \label{acr}
  \begin{split}
    \{a(f),a(g)\}&= \{a^*(f),a^*(g)\}=
    \{a(f),b(g)\}=\{a^*(f),b^*(g)\}\\
    &=\{a^*(f),b(g)\}=\{a(f),b^*(g)\}=0,
  \end{split} 
\end{equation}
and
\begin{equation}
\{a(f),a^*(g)\}=(f,\Pvp g), \, \{b^*(f),b(g)\}=(f,\Pvn g).
\end{equation}
Formally this is equivalent to 
\begin{equation}
  \label{facr1}
  \begin{split}
    \{a(x),a(y)\}&= \{a^*(x),a^*(y)\}= \{a(x),b(y)\}=\{a^*(x),b^*(y)\}\\
    &=\{a^*(x),b(y)\}=\{a(x),b^*(y)\}=0,
  \end{split}
\end{equation}
and
\begin{equation}
  \label{facr2}
  \{a(x),a^*(y)\}=\Pvp (x,y), \, \{b^*(x),b(y)\}=\Pvn(x,y)
\end{equation}
where $ \Pvp(x,y), \Pvn(x,y)$ are the integral kernels of the
projectors $ \Pvp, \Pvn$.

If there is no external potential there should be no polarization
effects present. It is therefore expected that the difference $\Qv$ of
one-particle density matrices of the perturbed and unperturbed vacua
\begin{equation}
  \label{eq:q}
  Q^\varphi:=\Pvp-\Pop= -\Pvn+\Pon
\end{equation}
plays a central role in defining the renormalized Hamiltonian. Using
Cauchy's formula we can express the $Q^\varphi$ in terms of the
respective resolvents (Kato \cite{Kato1966}, Section VI,5, Lemma 5.6)
  \begin{equation}
    \label{eq:12}
    Q^\varphi= \frac1{2\pi} \dK 
    \left(\frac1{D^\varphi+i\eta}-\frac1{D^0+i\eta}\right)
  \end{equation}
  The difference of $\Qv$ and of the first order resolvent expansion,
  i.e., 
\begin{equation}
  \label{eq:q2}
  \Qv-\frac\alpha{2\pi} \dK \ide \varphi\ide
\end{equation}
will contribute to the renormalized operator as follows: one interprets
its spin summed diagonal as density. The corresponding electric
potential should be added to the one particle operator. To avoid any
unnecessary difficulties defining the operator we split \eqref{eq:q2}
again in three summands motivated by iterating the resolvent
equation:
\begin{equation}
  \label{eq:20}
  \alpha^2Q_2 + \alpha^3 Q_3 +  \alpha^4Q_4,
\end{equation}
where the indices $2,3,4$ indicate the number of $\varphi$'s in the
expression, i.e.,
\begin{equation}
  \label{eq:13}
  \begin{split}
    Q_2 &:= \frac1{2\pi} \dK \ide \varphi\ide \varphi \ide,\\
    Q_3 &:= \frac1{2\pi} \dK \ide \varphi\ide \varphi 
    \ide \varphi \ide,\\
    Q_4 &: = \frac1{2\pi} \dK \ide \varphi\ide \varphi
    \idve \varphi \ide \varphi \ide.
    \end{split}
\end{equation}

We can immediately remark that the density corresponding to $Q_2$
vanishes: the terms linear in the Dirac matrices vanish after
summation over $\sigma$, since the Dirac matrices are traceless; the
remaining terms are odd in $\eta$ and vanish after integration over
$\eta$.  Consequently, we can disregard this term in defining the
operator.

We now define the density
\begin{equation}
  \label{eq:21}
  \rho_3(\bx)
  := (2\pi)^{-3}\int_{\rz^3} d\bp  \int_{\rz^3} d\bq \sum_{\sigma=1}^4 
  e^{i(\bp-\bq)\cdot\bx}\hat Q_3(\bp,\sigma;\bq,\sigma) 
\end{equation}
where
\begin{multline}
  \label{eq:23}
  \hat Q_3(\bp,\bq) = \frac1{2\pi}\dK\int_{\rz^3} d\bp_1 \int_{\rz^3}d\bp_2
  (D_\bp+i\eta)^{-1} \circ\hat\varphi(\bp-\bp_1)
  \circ(D_{\bp_1}+i\eta)^{-1}\\
  \circ \hat\varphi(\bp_1-\bp_2)\circ(D_{\bp_2}+i\eta)^{-1}
  \circ\hat\varphi(\bp_2-\bq)\circ (D_{\bq}+i\eta)^{-1}
\end{multline}
with $D_\br:= \alp\cdot\br+\beta$.  The corresponding electric potential
is
\begin{equation}
  \label{eq:P3}
   P_3 :=\rho_3*|\cdot|^{-1}.
\end{equation}
The quadratic form defining $P_4$ is given by
\begin{equation}
  \label{eq:22}
   (\psi,P_4\psi) := \tr (\chi Q_4)
\end{equation}
where $\chi(\bx):=\int dy |\psi(y)|^2/|\by-\bx|$.

It will be useful to introduce the function $C$ 
\begin{equation}
  \label{eq:C}
  \begin{split}
    C(\bk) &= \frac12 \bk^2 \int_0^1 dx (1-x^2)
    \log[1+\bk^2(1-x^2)/4] \\
    &= \frac13\bk^2 \left[(1- \frac 2{\bk^2})\sqrt{1+\frac4{\bk^2}}\log
      \frac{\sqrt{1 + 4/\bk^2} +1}{\sqrt{1 + 4/\bk^2}
        -1}+\frac4{\bk^2}-\frac53\right]
\end{split}
\end{equation}
as already done by Serber \cite{Serber1935} and Uehling
\cite{Uehling1935} and later by Pauli and Rose \cite{PauliRose1936},
Jauch and Rohrlich \cite{JauchRohrlich1955}, Schwinger
\cite{Schwinger1949}, and Klaus and Scharf \cite{KlausScharf1977V}.
The vacuum polarization potential $U$, also known as \textit{Uehling
  potential}, is defined via its Fourier transform
\begin{equation}
  \label{eq:vvac}
  \hat U(\bk) = \hat \varphi(\bk)
  \frac{C(\bk)}{\pi|\bk|^2}.
\end{equation}

The renormalized one-particle operator is
\begin{equation}
  \label{eq:dren}
  D_{\mathrm{ren}}:= \alp\cdot\frac1i\nabla+\beta - \alpha \varphi
  - \alpha^2 U +\alpha^2 X - \alpha^4 P_3 - \alpha^5 P_4
\end{equation}
where $X$ is the renormalized operator with integral kernel
\begin{equation}
  \label{eq:1}
  \alpha X(x,y):=\frac{\Qv(x,y)}{|\bx-\by|}.
\end{equation}

To introduce the operator $D_{\mathrm{ren}}$ might appear unmotivated
at this point. However, it has a solid physical motivation: it emerges
through mass and charge renormalization from the canonical formal text
book Hamiltonian (see, e.g., Milonni \cite{Milonni1994}, p. 385,
Formula (11.25)) for the interaction of electrons when there are no
photons present. In turn the charge and mass renormalization
originates in three physical principles \textbf{W1, W2,} and
\textbf{W3} as we will explain in Section \ref{physmot}.

Moreover, and this is our main mathematical result, the operator $\Pvp
D_{\mathrm{ren}}$ on $\Hp$ turns out to be well defined and
self-adjoint on the same form domain as the free Dirac operator:
\begin{theorem}
  \label{sec:model}
  Assume $\alpha Z\in[0,1)$ and $\varphi=n*|\cdot|^{-1}$ with
  spherically symmetric Schwartz function with compact support in
  Fourier space.  Then the quadratic forms 
  $(\psi,U\psi)$, $(\psi,X\psi)$, $(\psi,P_3\psi)$, $(\psi,P_4\psi)$
  on $\Pvp(H^1(\Gamma))$ are relatively form bounded with
  respect to $(\psi,|D^0|\psi)$ with form bound zero.
\end{theorem}
This has the following consequence:
\begin{corollary}
  \label{sec:model-1}
  There exists a unique self-adjoint operator $D_+$ fulfilling
  $(\psi,D_+\psi)=(\psi, (\dv +\alpha^2 X - \alpha^2 U - \alpha^4 P_3
  - \alpha^5 P_4)\psi)$ for all $\psi\in\Pvp(H^1(\Gamma))$ with
  form domain $\Pvp (H^{1/2}(\Gamma))$.  Furthermore it is bounded from
  below. 
\end{corollary}
\begin{proof}
  Define $c:=\sqrt2$ and $M:=\sqrt2\|\varphi\|_\infty$. Then,
  obviously $0\leq (c^2-2) (D^0)^2 - 2 \varphi^2+M^2$ implying
  $(D^0)^2 \leq c^2 (D_0^2 - 2 \Re(\varphi D^0)
  +\varphi^2)+M^2+2Mc|\dv|$. Rewriting this and taking the square root
  -- which is an operator montone function -- yields
  \begin{equation}
    \label{eq:24}
    |D^0|\leq c |\dv| + M.
  \end{equation}
  This means that infinitesimal form boundedness with respect to
  $|D^0|$ implies also infinitesimal form boundedness with respect to
  $|\dv|$. Moreover, we have $(\psi,|\dv|\psi)= (\psi,\dv\psi)$ for
  $\psi\in\Pvp(H^1(\Gamma))$.  
  
  Thus, according to the KLMN theorem (Reed and Simon
  \cite{ReedSimon1978}, Theorem X.17) there exists a unique
  self-adjoint operator $D_+$ on $\Hp$ whose form
  domain is the form domain of $\dv$ (which -- since $\varphi$ is
  bounded -- is the form domain of $\Pvp D^0$, i.e., $\Pvp
  H^{1/2}(\Gamma)$).  Moreover, the quadratic form of $D_+$ fulfills 
  \begin{equation}
    \label{eq:25}
    (\psi,D_+\psi)= (\psi, (\dv +\alpha^2 X - \alpha^2 U -
  \alpha^4 P_3 - \alpha^5 P_4)\psi).
  \end{equation}

\end{proof}
As already mentioned in the introduction, the Hamiltonian
(\ref{eq:dren}) can be used to describe $\mu$-mesonic atoms where the
interaction with the photon field is negligible as indicated
experimentally by Peterman and Yamaguchi \cite{PetermanYamaguchi1959}
and theoretically by Glauber et al \cite{Glauberetal1960}.

\section{Physical ``Derivation'' of the Renormalized
  Hamiltonian\label{physmot}} 

We start with the formal expression for the interaction of electrons
when no photons are present (Kroll and Lamb \cite{KrollLamb1949},
French and Weisskopf \cite{FrenchWeisskopf1949}) found also in text
books (see, e.g., Milonni \cite{Milonni1994}):
\begin{equation}
  \label{ie}
  \W_\mathrm{ur} 
  = \frac12 \int dx \int dy
  \frac{\Psi^*(x)\Psi(x)\Psi^*(y)\Psi(y)}{|\bx - \by|}. 
\end{equation}
The Hamiltonian describing our system is formally given by
\begin{equation}\label{hstart}
  \Hh_\mathrm{ur} = \mathbb{D}^\varphi_\mathrm{ur} +  \alpha\W_\mathrm{ur}
\end{equation}
where
\begin{equation}
  \label{eq:1part}
  \mathbb{D}^\varphi_\mathrm{ur}:=\int dx\, \Psi^*(x) 
  D^\varphi\Psi(x).
\end{equation}
It is well known that this expression contains several singular terms.
In particular it does not even contain any normal ordering. The
remaining part of this section can be viewed as manipulating on it a
physical allowed way and transforming it to a physically equivalent
expression that is mathematically meaningful, namely the renormalized
Hamiltonian. We emphasize that none of the steps taken is
mathematically justified, i.e., the eventual justification of the
renormalized Hamiltonian is its successful predictive power.

We use three guiding principles to transform expressions for the
energy into other physically equivalent ones as formulated and
justified by Weisskopf \cite{Weisskopf1936}, p.  6: \textit{``The following
three properties of the vacuum electrons are assumed to be irrelevant:}
\begin{description}
\item[\textbf{W1}] \textit{The energy of the vacuum electrons in
    field free space.}\label{w1}
\item[\textbf{W2}] \label{w2} \textit{The charge and current density of the
    vacuum electrons in field free space.}
\item[\textbf{W3}] \textit{A field independent electric and magnetic
    polarizability that is constant in space and time.''}\label{w3}
\end{description}

Similar procedures have been suggested by Heisenberg
\cite{Heisenberg1934}, French and Weisskopf
\cite{FrenchWeisskopf1949}, Kroll and Lamb \cite{KrollLamb1949}, and
Dyson \cite{Dyson1949}.

Exploiting the canonical anti-commutation relations (\ref{facr1})
we can rewrite (\ref{hstart}). For the one-particle part we have
\begin{equation}
  \label{eq:fie}
  \mathbb{D}^\varphi_\mathrm{ur} 
  = \mathbb{D}^\varphi + \int dx (\dv P^\varphi_-)(x,x).
\end{equation}
The last summand is a -- although infinite -- constant which we drop,
since it does not influence energy differences.  For the two-particle
part we get
\begin{equation}\label{fie}
  \begin{split}
    \W_\mathrm{ur} =& \frac12 \int dx\int dy 
    \frac{:\Psi^*(x)\Psi(x)\Psi^*(y)\Psi(y):}{|\bx-\by|}
    \\
    &+ \frac12 \int dx\int dy : \Psi^*(x)\Psi(y): \frac{\Pvp (x,y) -
      \Pvn(x,y)}{|\bx-\by|} \\
    &+ \int dx\int dy : \Psi^*(x)\Psi(x): \frac{\Pvn(y,y)}{|\bx-\by|} \\
    &+  \frac12 \int dx\int dy \frac{ \Pvp (x,y) \Pvn(x,y)}{|\bx-\by|} +
    \frac12 \int dx\int dy \frac{ \Pvn (x,x) \Pvn(y,y)}{|\bx-\by|}.
  \end{split}
\end{equation}
The last two terms are again constants which we drop. The first term
is the normal ordered two-particle interaction  which has finite
expectation in states of finite kinetic energy. We will denote it by
\begin{equation}
  \label{eq:w}
  \W= \frac12 \int dx\int dy 
  \frac{:\Psi^*(x)\Psi(x)\Psi^*(y)\Psi(y):}{|\bx-\by|}.
\end{equation}
The remaining two other terms are one-particle operators of particular
interest to us.

Both terms, the classical electrostatic interaction energy of the electron
with the polarized Dirac sea called the ``non-exchange energy''
\begin{equation}\label{pur}
  \mathbb{P}_\mathrm{ur} := - \int dx\int dy  
  : \Psi^*(x)\Psi(x): \frac{\Pvn(y,y)}{|\bx-\by|},
\end{equation}
and the exchange energy
\begin{equation}\label{xur}
  \mathbb{X}_\mathrm{ur} := \frac12 \int dx\int dy 
  : \Psi^*(x)\Psi(y): \frac{\Pvp (x,y) - \Pvn(x,y)}{|\bx-\by|}
\end{equation} 
are not well defined. (For curiosity we remark that the latter is
logarithmically divergent in $\Lambda$, if one introduces a cut-off by
$\Psi(x) = \int_{|\bp| \leq \Lambda} \hat \Psi (\bp,\sigma) e^{-i \bp\cdot
  \bx} d\bp$).

To renormalize the exchange energy we introduce the operators $\Pop,
\Pon$ which are the projectors on the positive and negative subspace
of the free Dirac operator $ D^0$.  (Note that we can interpret
$\Pon$ as the one-particle density matrix of the free Dirac sea.)

\subsection{The Renormalization of the Exchange Energy\label{sxren}}

To renormalize $\mathbb{X}_\mathrm{ur}$ we subtract the exchange
interaction energy of the electron with the free Dirac sea using
Principle \textbf{W1}, i.e.,
\begin{equation}
  \label{x}
  \begin{split}
    \alpha\mathbb{X} :=&\frac12 \int dx\int dy : \Psi^*(x)\Psi(y):
    \frac{[(\Pvp
      -\Pop) - (\Pvn - \Pon)](x,y)}{|\bx-\by|}\\
    =& \alpha\int dx\int dy : \Psi^*(x)\Psi(y):X(x,y)
  \end{split}
\end{equation}
with $X$  as defined in \eqref{eq:1}.

(In physical language this subtraction of an undefined operator --
known as ``counter term'' -- is called ``mass renormalization''. We
refer to French and Weisskopf \cite{FrenchWeisskopf1949}, Equation
(30), for the motivation of this terminology.)

From now on, we will assume that the external potential $\varphi$ is
so weak that all there are only positive eigenvalues in the gap of
$\dv$. Then,
\begin{equation}
 \Pvp =\frac 12 +  \frac 1{2\pi}\int_{-\infty}^{\infty}d\eta \idve
\end{equation}
as well as
\begin{equation}
 \Pop =\frac 12 +  \frac 1{2\pi}\int_{-\infty}^{\infty}d\eta \ide
\end{equation}
(Kato \cite{Kato1966}, Section VI,5, Lemma 5.6). Thus,
\begin{equation}\label{QV}
  \Qv= \frac{\alpha}{2\pi} \dK\ide
  \varphi\ide +\frac{\alpha^2}{2\pi} \dK\ide \varphi\idve
  \varphi\ide
\end{equation}
where the first summand of the right hand side is denoted by $\alpha
Q_1$ and the second summand is treated in \eqref{eq:q2} through
\eqref{eq:13}.

Furthermore, since
$$\int_{\rz^3} d\bp \frac{\bp^2|\hat\varphi(\bp)|^2}{1+|\bp|}<\infty$$
the potential $\varphi$ is regular in the sense of Klaus and Scharf
\cite{KlausScharf1977T}, namely the operator $\Qv\in
\mathfrak{S_2}(\mathfrak{H})$, i.e., $\Qv$ is an Hilbert-Schmidt
operator. (See also \cite{HelfferSiedentop1998}, Theorem 4.)  This
allows to show the finiteness of the exchange energy between the
one-particle density matrix of the electron-positron field and the
difference of the density matrices of the polarized Dirac sea and the
free Dirac sea.

To formulate the next lemma we fix the following notation: let
$C_{p,q}$ be the optimal constant in the generalized Young inequality,
i.e., $\|f*g\|_r\leq C_{p,q}\|f\|_p \|g\|_q$, $1<p,q,r<\infty$,
$r^{-1}+1=p^{-1}+q^{-1}$. 
\begin{lemma}\label{lem1}
  Let $\psi \in L^3(\Gamma)\cap L^2(\Gamma)$.  Then
  \begin{equation}\label{eq:g1}
    \left|\int dx\int dy
      \frac{\overline{\psi(x)}\Qv(x,y)\psi(y)}{|\bx-\by|}\right| 
    \leq \sqrt{C_{3/2,3/2}\|1/|\cdot|^2\|_{3/2,w}} \|\Qv\|_2 \|\psi\|^2_3,
  \end{equation}
  and for every $\epsilon >0$ there exists a
  constant $C_\epsilon>0$ such that
  \begin{equation}\label{eq:g2}
    \left|\int dx\int dy 
      \frac{\overline{\psi(x)}\Qv(x,y)\psi(y)}{|\bx-\by|}\right|
    \leq \epsilon\|\psi\|^2_3 + C_\epsilon\|\psi\|_2^2.
  \end{equation}
\end{lemma}
\begin{proof}
  Since $\Qv $ is a Hilbert-Schmidt operator we get using the Schwarz
  inequality
  \begin{multline}\label{223}
    L:=\left|\int dx\int dy \frac{\overline{\psi(x)}
        \Qv(x,y)\psi(y)}{|\bx-\by|}\right|\\
    \leq \left(\int dx \int dy
      \frac{|\psi(x)|^2|\psi(y)|^2}{|\bx-\by|^2}\right)^{1/2}
    \left(\int dx\int dy |\Qv(x,y)|^2\right)^{1/2}.
  \end{multline}
  The second factor of the right hand side is the Hilbert-Schmidt norm
  $\|\Qv\|_2$ of $\Qv$.
  
  To estimate the first factor we decompose the kernel into two
  functions $f(\bx):= \chi_{B_R(0)}(\bx)/|\bx|^2$ and the rest $g$,
  i.e., $1/|\bx|^2 = f(\bx)+g(\bx)$.
  
  Thus, using inequality \eqref{223} we get
  \begin{equation}\label{eq:16}
    L\leq \left[(|\psi|^2 * f,|\psi|^2)^{1/2}+ (|\psi|^2
    *g,|\psi|^2)^{1/2}\right] \|\Qv\|_2.
  \end{equation}
  We estimate the first and second summand of the first factor on the
  right hand side separately from above.
  
  The first summand yields using the H\"older inequality followed by
  the generalized Young inequality (see, e.g., Reed and Simon
  \cite{ReedSimon1975}, p. 32)
\begin{equation}
  \label{h1}
  (|\psi|^2 *f,|\psi|^2) 
\leq C_{3/2,3/2}\|\psi^2\|_{3/2}^2\|f\|_{3/2, w},
\end{equation}
where $w$ indicates the weak-norm. 

Picking the radius $R=\infty$, i.e., $g=0$, yields immediately
\eqref{eq:g1}.

To prove \eqref{eq:g2} we also use \eqref{h1} but pick the radius
$R>0$ sufficiently small: in this case we need to bound also the
second summand containing $g$; we use again H\"older's inequality now
followed by using Young's inequality
\begin{equation}
  \label{h2}
  (|\psi|^2 *g,|\psi|^2)   \leq \|\psi^2\|_1^2 \|g\|_{\infty}.
\end{equation}
Thus, the first factor on the right hand side of \eqref{eq:16} is
bounded by
$$\sqrt {C_{3/2,3/2}}\|\psi^2\|_{3/2}\|f\|^{1/2}_{3/2, w}+ \|\psi^2\|_1
\|g\|_\infty^{1/2}.$$
Since $\|f\|_{3/2, w}$ tends to zero as $R$ tends to
zero, the claimed inequality follows.  \end{proof}

\subsection{Electrostatic Vacuum Polarization Energy 
  (Non-Exchange Energy)\label{srennonx}}

In the expression for the electrostatic vacuum polarization energy we
replace the density of the polarized sea by the difference of this
density and the free Dirac sea using Principle \textbf{W2}:
\begin{equation}
  \label{P}
  \begin{split}
    \tilde{\mathbb{P}} = &-\int dx\int dy : \Psi^*(x)\Psi(x):
    \frac{\Pvn(y,y)-\Pon(y,y)}{|\bx-\by|} \\
    =&  \int dx\int dy : \Psi^*(x)\Psi(x):
    \frac{\Qv(y,y)}{|\bx-\by|} \\
    =& \int dx\int_{\rz^3} d\by : \Psi^*(x)\Psi(x): \frac{\Tr
      \Qv(\by,\by)}{|\bx-\by|}.
  \end{split}
\end{equation}
(Here and in the following we will denote by $\Qv(\bx,\by)$ the
$4\times4$ matrix with entries
$\left(\Qv(\bx,\sigma;\by,\tau)\right)_{\sigma,\tau=1}^4$.) However,
the integral kernel of $\Qv$ is always singular on the diagonal except
for vanishing potential as can be seen from \eqref{logp} implying that
$\tilde{\mathbb{P}}$ is not well defined; one more renormalization is
necessary. The question how to extract the physical relevant
information from $\Tr \Qv(\by,\by)$ was already asked by Dirac
\cite{Dirac1934} and partially answered by Dirac \cite{Dirac1934D},
Heisenberg \cite{Heisenberg1934}, Serber \cite{Serber1935}, Uehling
\cite{Uehling1935}, Weisskopf \cite{Weisskopf1936}, Schwinger
\cite{Schwinger1949}, Dyson \cite{Dyson1949}, Klaus and Scharf
\cite{KlausScharf1977V}, and others. The proposed solution amounted to
a perturbative renormalization according to Principle \textbf{W3}. ---
One of our main results is that this renormalization can be done
non-perturbatively: subtracting the zeroth order expansion $Q_1$ of
the difference $\Qv$ of $\Pvp$ and $\Pop$ will turn $\Tr \Qv(\by,\by)$
into well defined quantities given in \eqref{eq:P3} and \eqref{eq:22}.

Recall that
\begin{equation}
  Q_1 = \frac1{2\pi} \dK \ide \varphi \ide.
\end{equation}
Thus, in momentum space $Q_1$ is given by
\begin{equation}
\hat Q_1(\bp,\bq) = (2\pi)^{-5/2} \dK  
\frac{\alp\cdot\bp+\beta -i\eta}{\bp^2 +1 + \eta^2} \hat\varphi(\bp-\bq)\frac{\alp \cdot \bq + \beta -i\eta}{\bq^2 +1 +\eta^2},
\end{equation}
which leads to 
\begin{equation}
\Tr \hat Q_1(\bp,\bq) = 2^{-1/2}\pi^{-3/2} \hat\varphi(\bp-\bq)
\frac{\bp \cdot \bq + 1-E(\bp)E(\bq)}{E(\bp)E(\bq)(E(\bp) + E(\bq))}
\end{equation}
by a straightforward calculation with $ E(\bp) = \sqrt{\bp^2 +1}$.  In
configuration space we obtain
\begin{equation}
  \begin{split}
    \Tr Q_1(\bx,\by) &= (2\pi)^{-3}\int_{\rz^3} d\br \int_{\rz^3} d\bq
   e^{i\br\cdot \bx } \Tr\hat Q_1(\br,\bq) e^{-i\bq\cdot \by } \\
    &= (2\pi)^{-3}\int d\bp d\bk \Tr \hat Q_1(\bp-\bk/2,\bp + \bk/2) 
    e^{i\bp \cdot (\bx-\by)} e^{-i\bk\cdot (\bx + \by)/2} \\
    &=: \tq \big(\bx-\by,\frac{\bx+\by}2\big)
    \end{split}
\end{equation}
after introducing new variables of integration $\br=\bp-\bk/2$ and
$\bq=\bp+\bk/2$.  Defining $\boldsymbol{\xi} := \bx - \by$ we remark
that the ``limits'' $ \lim_{\by \to \bx} \Tr Q_1(\bx,\by)$ and
$\lim_{\xi \to 0} \tq(\xi,\bx)$ are formally the same.  The
corresponding expression $\tilde{\mathbb{P}}$ in the electrostatic
energy \eqref{P} becomes formally
\begin{equation}
  \label{eq:6}
  \int dx \int d\by \frac{:\Psi^*(x)\Psi(x): \Tr Q_1(\by,\by)}{|\bx-\by|} = 
  \int dk \widehat { : \Psi^*\Psi:}(k) \frac{4\pi}{ \bk^2} \hat \tq (0,\bk),
\end{equation}
where $\hat \tq(\xi,\cdot)$ is the Fourier transform of $\tq$ with
respect to the second variable for fixed $\xi\neq0$, i.e., formally
\begin{multline}
  \label{logp}
  \hat \tq(\xi,\bk)= (2\pi)^{-3/2} \int_{\rz^3} d\bp\ \Tr \hat
  Q_1(\bp-\bk/2,\bp + \bk/2) e^{i \bp \cdot
    \boldsymbol{\xi}}\\
  = \frac1{4\pi^3} \hat \varphi(\bk)\int_{\rz^3} d\bp\frac{\bp^2 - \bk^2/4
    +1 - E(\bp-\bk/2) E(\bp +\bk/2)} { E(\bp -\bk/2) E(\bp +\bk/2)\left(
    E(\bp-\bk/2) + E(\bp+\bk/2)\right)} e^{i \bp \cdot \boldsymbol{\xi}}.
\end{multline}

We note that the integral (\ref{logp}) is logarithmically divergent at
$\boldsymbol\xi=0$ independently of the form of the external potential
$\varphi$.  This shows -- as already remarked above -- that the limit
$\lim_{\by \to \bx} \Tr \Qv (\bx,\by)$ only exists, if $\varphi$
vanishes.

The expression $\hat \tq(\boldsymbol\xi,\bk)$ has been intensively
studied in the literature (see, e.g., Heisenberg
\cite{Heisenberg1934}, Serber \cite{Serber1935}, Pauli and Rose
\cite{PauliRose1936}, Weisskopf \cite{Weisskopf1936}, and Klaus and
Scharf \cite{KlausScharf1977V}).  We will follow mainly the
calculations of Pauli and Rose.  (Since their treatment is
time-dependent one has to set $k_0 = 0$ to translate to our
situation.)  According to \cite{PauliRose1936}, Equations (5) -- (9),
we can separate $\hat \tq$ into two terms
\begin{equation}
  \label{decom}
  \hat \tq (\boldsymbol\xi,\bk) 
  = F_1(\boldsymbol\xi,\bk) +  \hat \varphi(\bk) \bk^2 F_0(\boldsymbol\xi),
\end{equation}
with 
\begin{equation}\label{fxi}
  F_0(\boldsymbol\xi)= -\frac 1{16\pi^3} 
  \int_{\rz^3} d\bp \left( 1- \frac{\bp^2 \cos^2\theta}{1 +
      \bp^2}\right)\frac{e^{i\bp \cdot \boldsymbol\xi}}{(1 + \bp^2)^{3/2}}
\end{equation}
$\theta$ being the angle between $\boldsymbol\xi$ and $\bp$. With this
definition of $F_0$ the function 
$F_1$ is finite for $\boldsymbol\xi = 0$ and has there the value
\begin{multline}\label{rhov}
  \hat \rho_\mathrm{vac} (\bk) := F_1 (0,\bk) \\
  = \frac{\hat \varphi(\bk)}{4\pi^3}\int_{\rz^3} \frac{\bp^2
      - \frac{\bk^2}4 +1 - E(\bp -\frac\bk2) E(\bp +\frac\bk2)} {E(\bp
      -\frac\bk2) E(\bp +\frac\bk2)\left( E(\bp -\frac\bk2) + E(\bp
        +\frac\bk2)\right)} + \bk^2
    \frac{\bp^2\sin^2\theta+1}{4E(\bp)^5} d\bp\\
  = \frac1{4\pi^2}\hat \varphi(\bk) C(\bk)
\end{multline}
where $C$ is the function defined in \eqref{eq:C}.  While each of the
summands in the latter formula decreases like $|\bp|^{-3}$ for large
values of $|\bp|$ and therefore the corresponding parts of the integral
are logarithmically divergent, the difference in the integrand
decreases like $|\bp|^{-5}$ and is therefore convergent.

Pauli and Rose \cite{PauliRose1936} obtain the following asymptotic
behavior for $C$:
\begin{equation}
  \label{cdk}
  C(\bk)/\bk^2 = 
  \begin{cases}
    \frac1{15} \bk^2 + o(\bk^2)& |\bk|\rightarrow 0\\
    \frac 23 \log(|\bk|) - \frac 59 + o(1)& |\bk|\to\infty
  \end{cases}.
\end{equation}

We note that the second summand on the right hand side of
\eqref{decom} can be written as $4\pi\hat
n(\bk)F_0(\boldsymbol{\xi})$, i.e., this depends only on the density
of the nucleus. This implies that it can be dropped according to
\textbf{W3}.  This means that $\rho_\mathrm{vac}$ as defined in
\eqref{rhov} can be considered as the physically relevant density of
the polarized vacuum.

\subsection{The Fully Renormalized Hamiltonian}
\label{sec:fully-renorm-hamilt}
The potential of the vacuum polarization density is
\begin{equation}
  \label{eq:7}
  U  = |\cdot|^{-1}*\rho_\mathrm{vac}; 
\end{equation}
the corresponding term of the electrostatic interaction of the
electron-pos\-i\-tron field with the vacuum reads
\begin{multline}
  \label{eq:4}
  - \alpha^2\int dx : \Psi^*(x) \Psi(x): U (\bx)
  = -\alpha^2\int dx :\Psi^*(x) \Psi(x): |\cdot|^{-1}*\rho_\mathrm{vac}(\bx)\\
  =- \alpha^2\int dk \widehat{:\Psi^* \Psi:}(k) \frac{4\pi}{
    \bk^2}\rho_\mathrm{vac}(\bk)
\end{multline}
where we used $\mathcal{F}(|\cdot|^{-1}) = \sqrt{2/\pi} |\cdot|^{-2}$.

According to (\ref{P}), \eqref{QV}, \eqref{eq:21}, and \eqref{eq:P3}
the second quantized renormalized polarization energy becomes
\begin{equation}
  \label{wren}
  \mathbb{P} =\int dx : \Psi^*(x) \Psi(x):
  \left(U (\bx) + \alpha^2P_3(\bx)+ \alpha^3P_4(\bx)\right).
\end{equation}
Consequently our fully renormalized Hamiltonian is
\begin{equation}
  \label{hren}
  \boxed{\Hh = \mathbb{D}^\mathrm{\varphi} +\alpha \mathbb{W} 
+ \alpha^2\left(-\mathbb{P}    + \mathbb{X}\right)}.
\end{equation}

\subsection{Physical Interpretation of the Renormalization Procedure
  of the Vacuum Polarization}
\label{sec:phys-interpr-renorm}

In physics literature the subtraction of the singular part of the
diagonal term of the one-particle density matrix of the Dirac sea,
i.e., the dropping of the second summand \eqref{decom}) is called
\textit{charge renormalization} for the following reason: the term
subtracted from $\tilde{\mathbb{P}}$ to obtain $\mathbb{P}$ is
\begin{equation}
  \label{eq:5}
  \alpha^2 4\pi F_0(\boldsymbol\xi)
  \int dx :\Psi^*(x) \Psi(x): \varphi (\bx).
\end{equation}
Formally this result can also be obtained by replacing the square of
the (bare) charge $e^2=\alpha$ in $\dv$
\begin{equation}
  \label{obsch}
  e^2 \mapsto e^2(1-4\pi e^2 F_0(\boldsymbol\xi)).
\end{equation}
Note, that
$$F_0(\boldsymbol\xi) = - \log(|\boldsymbol\xi|) + o(1) $$
for small
$|\boldsymbol\xi|$ which leads to a well known formula in the
literature (see, e.g., Milonni \cite{Milonni1994}, p. 417).  It is
interesting that a change in the effective charge due to the
polarization of the vacuum was already suggested by Furry and
Oppenheimer \cite{FurryOppenheimer1934}.

The word {\it polarization} is due to the following picture: according
to Dirac the electrostatic field causes a redistribution of charge in
the Dirac sea, i.e., it \textit{polarizes} the vacuum.  In particular
the nucleus polarizes the vacuum in its vicinity causing a screening
lowering the effective charge for an observer at a distance.

Another reason for the fact that the infinity of the diagonal part of
the density matrix of the Dirac sea in (\ref{decom}) is invisible in
experiments is the following: in first order the factor in front of
$F_0(0)$ changes $\alpha$ in $\dv$ by an (infinite) constant only,
which does not effect the degeneracy of the eigenvalues of $\dv$,
i.e., it does not cause a splitting of degenerated eigenvalues, a fact
that is confirmed by experiment.

\subsection{The Vacuum Polarization Potential of the Coulomb 
  Potential\label{coulomb}}
Recall that the nuclear potential is $\varphi=|\cdot|^{-1}*n$;
consequently $\hat \varphi(\bk) = 4\pi\hat n(\bk)/\bk^2$.  Thus the
Fourier transform of the vacuum polarization potential using
(\ref{rhov}) gives
\begin{equation}
  \label{vvacz}
  \hat U (\bk)= \frac{\hat \varphi(\bk)C(\bk)}{\pi|\bk|^2} 
  = 4\frac{\hat n(\bk)C(\bk)}{|\bk|^4}
\end{equation}
which is spherically symmetric and compactly supported under our
assumptions on the charge distributions of the nucleus. 

If we assume that the nuclear density is a spherically symmetric Schwartz
function -- as we do in the mathematical part of this paper -- this
implies that the vacuum polarization potential $U$ is bounded
continuous and decreases exponentially at infinity.

However, if we assume that we have a point nucleus as assumed by
Uehling \cite{Uehling1935}, this is no longer the case. To relate to
Uehling's work we will discuss this case as well although the
corresponding potential is no longer form bounded with respect to the
kinetic energy. 

From (\ref{vvacz}) we have
\begin{equation}
  \label{eq:9}
  U = Z \sqrt{\frac2{\pi^3}}\mathcal{F}^{-1}\left(\frac C{|\cdot|^4}\right).
\end{equation}
According to Uehling \cite{Uehling1935} and Schwinger
\cite{Schwinger1949}, Equation (2.53), this is
\begin{equation}
  \label{eq:11}
  U(\bx) = \frac2{3\pi}\frac Z{|\bx|} \int_1^\infty e^{-2|\bx|s} 
  \big(1 + \frac 1{2s^2}\big)\frac{(s^2 -1)^{1/2}}{s^2} ds,
\end{equation}
which means asymptotically
\begin{equation}
  \label{eq:10}
  U(\bx) = 
  \begin{cases}
    -\frac2{3\pi}Z|\bx|^{-1} \left(\log|\bx| + \frac 56+\gamma\right) +
    O(1) &
    |\bx|\to 0\\
    \frac{Z}{4\sqrt\pi} e^{-2|\bx|}|\bx|^{-5/2}(1 + O(1/|\bx|)) & |\bx|\to
    \infty
  \end{cases}
\end{equation}
where $\gamma\approx 0.5772$ is Euler's constant.  Consequently one
obtains an effective potential
\begin{equation}
  \label{ux}
  \varphi_\mathrm{eff} (\bx) = -\alpha \varphi -\alpha^2 U = -\alpha
  \frac Z{|\bx|} + \alpha^2 \frac2{3\pi}\frac Z{|\bx|} \left(\log|\bx| +
  \frac 56+\gamma\right) + O(1)
\end{equation}
close to the nucleus.

Obviously, due to vacuum polarization, the effective potential becomes more
singular than the Coulomb potential. This implies that 
the energy 
$$|\bp| - \alpha^2 \frac{2Z}{3\pi|\bx|} \log \frac 1{|\bx|}$$
is unbounded from
below for all positive values of $\alpha$ and $ Z$ and (\ref{ux}) is
no longer relatively form bounded with respect to the relativistic kinetic
energy operator.  This suggests to avoid the mathematical idealization
of a point nucleus and to take the experimental fact that the nuclei
are extended into account.

\subsection{Splitting of the Bound State Energies}
\label{sec:splitt-bound-state}

The effect of the vacuum polarization potential to lowest order in
$\alpha$ is given by the effective one-particle operator
$$\Pvp\left(\dv - \alpha^2 U(\bx)\right)\Pvp.$$
Therefore, the energy eigenvalues are shifted in lowest order of
$\alpha$ by
\begin{equation}
  \delta E = -\alpha^2 \int_{\rz^3} d\bx U(\bx) |\psi(\bx)|^2
\end{equation}
where $\psi(x)$ denotes an eigenstate of $\dv$.

To get a rough heuristic estimate on the numerical effect Uehling
\cite{Uehling1935} assumes the nucleus to be a point particle, i.e.,
its density is $n(\bx)=Z\delta(\bx)$, and takes the corresponding
Schr\"odinger eigenstates $\psi_{n,l}$ where $n$ is the principal
quantum number and $l$ the orbital-angular-momentum quantum number:
\begin{multline}
  \label{eq:8}  
  \delta E_{n,l} = -\alpha^2 \int_{\rz^3} d\bx U(\bx)
  |\psi_{n,l}(\bx)|^2 \approx -\frac{4Z\alpha^2}{15}|\psi_{n,l}(0)|^2 =
  - \frac{4Z^4\alpha^5}{15\pi n^3}\delta_{0,l}.
\end{multline}

Concerning the first excited eigenvalue $n=2$ this indicates an energy
level splitting (of the $2s$ and $2p$ state) of
\begin{equation}
  \label{levelsplit}
  \delta E_{2,0} \sim - \frac{Z^4 \alpha^5 m}{30},
\end{equation}
the \textit{Uehling effect} \cite{Uehling1935}, Equation (29).

The vacuum polarization (\ref{levelsplit}) accounts for only one
percent of the $2s_{1/2} - 2p_{1/2}$ Lamb shift of hydrogen, since the
Bohr radius is much bigger than the range of the vacuum polarization
potential. However, the Bohr radii of muonic atoms are much smaller
because of the large effective mass which means that the vacuum
polarization of muonic helium accounts for $90$ percent of the Lamb
shift (Peterman and Yamaguchi \cite{PetermanYamaguchi1959}, Glauber et
al. \cite{Glauberetal1960}, see also Greiner et al.
\cite{Greineretal1985}, p. 413).

\section{Proof of the Self-Adjointness of the Renormalized 
  Hamiltonian\label{mathres}}

In the following we are going to show that the quadratic form
associated to the operator $\Hh$ of the electron-positron field
restricted to the one-electron sector $\Hp$ defines a self-adjoint
operator which is bounded from below.  This means -- among other
things -- that higher order renormalizations, as introduced in
perturbation theory by Dyson \cite{Dyson1949}, are unnecessary.

We will show infinitesimal form boundedness of all perturbations,
namely $U$, $P_3$, and $P_4$ relative to $ \Pvp \dv \Pvp$. For $X$
this has been already shown in Lemma \ref{lem1}.

$\mathbf{U}$: According to \eqref{vvacz} and the remark thereafter $U$
is bounded and therefore trivially infinitesimally relatively form
bounded.

$\mathbf{P}_4$: We will show that for any positive $\epsilon$ there
exists a $C_\epsilon$ such that for all $\psi\in \mathcal{D}\big (
\dv\big) \subset \mathrm{L}^3(\Gamma) \cap\mathrm{L}^2(\Gamma)$
\begin{equation}
(\psi,P_4\psi) \leq \epsilon
\|\psi\|^2_3 + C_\epsilon\|\psi\|_2
\end{equation}
which implies the infinitesimal form boundedness by the Sobolev's
inequality for $\sqrt{-\Delta}$.

\begin{lemma}
  \label{lemq3}
  Assume $\alpha Z\in[0,1)$, $\chi\in L^5(\Gamma)$.  Then
\begin{equation}
   |\tr (\chi Q_4)| \leq \|\chi Q_4\|_1 
   \leq \frac{C_{\varphi,4}}{3\pi^2}
   \|\varphi\|_5^4\|\chi\|_5
\end{equation}
with
\begin{equation}
C_{\varphi,4}:=1 +\alpha \|\varphi\frac1{D^\varphi}\|.
\end{equation}
\end{lemma}
\begin{proof}
  We have
\begin{equation}
  \label{v4}
  \begin{split}
   |\tr (\chi Q_4)|
  \leq &\frac1{2\pi} \dK\left\| \chi\ide \varphi\ide
    \varphi \idve \varphi \ide \varphi
    \ide\right\|_1\\
  \leq &\frac1{2\pi} \dK \left\| \chi \ide\right\|_5\left\| \varphi \ide
  \right\|_5^3\left\|\varphi \idve\right\|_5.
  \end{split}
\end{equation}
(We use the standard notation $\| A\|_p = \sqrt[p]{\tr |A|^p}$.)  We
will estimate the right hand side of the above inequality which will
also show that $\chi Q^{(4)}\in \gS_1(\gH)$. To this end we estimate
the factor containing the perturbed resolvent:
\begin{multline}\label{eq:15}
  \|\varphi\idve \|_5 = \|\varphi \ide (D^0 + i\eta)\idve \|_5  \\
  \leq \| (D^0 + i\eta) \idve\|_\infty \| \varphi \ide \|_5.
\end{multline}
The first factor on the right side is finite:
$$
\| (D^0 + i\eta) \idve \|_\infty \leq 1 +\alpha \|\varphi \idve\|_\infty
\leq 1 +\alpha \|\varphi\frac1{D^\varphi}\|_\infty <\infty $$
where we use
that $D^\varphi$ is invertible because of Lemma \ref{DiracV} and that
$\varphi$ is bounded (see \eqref{eq:17}). (Note the boundedness would
also hold if $\varphi =Z\alpha/|\cdot|$, since $1/|\cdot|$ is
relatively bounded with respect to $\sqrt{-\Delta}$.)

Since  
\begin{equation}
\label{eq:18}
\|f(x)g(-i\nabla)\|_5 \leq (2\pi)^{-3/5} 
\|f\|_5 \|g\|_5,
\end{equation}
the norm being the one of the trace ideal $\gS_5(L^2(\rz^3))$ (Simon
\cite{Simon1979T}, Theorem 4.1), we can estimate the other factors
occurring in \eqref{v4}:
\begin{equation}
  \label{eq:14}
  \|\chi\ide\|_5\leq \frac{1}{2^{1/5}\pi^{3/5}}
\|\chi\|_5\|1/\sqrt{|\cdot|^2+1+\eta^2}\|_5
\end{equation}
(norm in $\gS_5(\gH)$ on the left hand side and in $\gS_5(L^2(\rz^3))$
on the right hand side) and a similar expression of the term
containing $\varphi$.

Using \eqref{eq:15} and \eqref{eq:14} allows us to continue \eqref{v4} as
\[
\begin{split}
  |\tr (\chi Q_4|\leq &\frac{C_{\varphi,4}}{4\pi^4 }\|\chi\|_5
  \|\varphi\|_5^4 \dK \int_{\rz^3}d\bp\frac1{(\bp^2 + 1 +
    \eta^2)^{5/2}}\\
  \leq &\frac{C_{\varphi,4}}{4\pi^4} \dK \frac 1{(\eta^2 +
    1)} \int_{\rz^3} \frac 1{(1+\bp^2)^2} d \bp\|\chi\|_5 \|\varphi\|_5^4
  \leq \frac{C_{\varphi,4}}{3\pi^2}\|\chi\|_5
  \|\varphi\|_5^4.
\end{split}
\]
\end{proof}
This has the 
\begin{corollary}
  \label{sec:proof-self-adjo}
  The perturbation $P_4$ is relatively form bounded with respect to $|D^0|$
  with form bound zero.
\end{corollary}
\begin{proof}
  We pick $\chi=|\psi|^2*|\cdot|^{-1}$ in Lemma \ref{lemq3} with
  $\psi\in H^{1/2}(\Gamma)$. Using Young's inequality followed by
  Sobolev's inequality yields the desired result.
\end{proof}
$\mathbf{P}_3$: Unfortunately, Simon's elegant trace inequality used
in (\ref{v4}) does not suffice to handle the $Q_3$ containing only
four resolvents. In that case we estimate directly:
\begin{lemma}\label{lemq2}
  Denote the electric potential of a state $\psi\in H^{1/2}(\Gamma)$
  of finite kinetic energy
  by $\chi(\bx):=\int dy |\psi(y)|^2/|\bx-\by|$ and assume $p>3$.
  Then there exists a constant $C_{\varphi,p}$ such that
  \begin{equation}
    |(\psi,P_3\psi)|\leq C_{\varphi,p} \|\chi\|_p.
  \end{equation}
\end{lemma}

\begin{proof}
We have
\begin{multline}\label{v3}
  (\psi,P_3\psi)=\int_{\rz^3} d\bx \chi(\bx)\rho_3(\bx) =\int_{\rz^3} d\bp
  \hat \chi(\bp)\widehat{\rho_3}(\bp) \\
  =(2\pi)^{-3/2} \int_{\rz^3}
  d\bp_1\int_{\rz^3} d\bp_2 \sum_{\sigma=1}^4 \hat\chi(\bp_1-\bp_2) 
  \hat Q_3(\bp_1,\sigma;\bp_2,\sigma)
\end{multline}
where we use the Definition \eqref{eq:21} of $\rho_3$.  The
``eigenfunctions'' of the free Dirac operator in momentum space are
\begin{equation}
  \label{eq:19}
  u_\tau(\bp):=
  \begin{cases}
    \frac 1 {N_+(\bp)} \begin{pmatrix} \sigma \cdot \bp {\bf e}_{\tau}\\
      -(1-E(\bp))\mathbf{e}_\tau\end{pmatrix} & \tau = 1,2,\\
    \frac 1 {N_-(\bp)} \begin{pmatrix} \sigma \cdot \bp \, {\bf e}_{\tau}\\
      -(1+E(\bp))\mathbf{e}_\tau\end{pmatrix} &\tau =3,4
  \end{cases}
\end{equation}
with $ {\bf e}_{\tau} := (1,0)^t$ for $\tau=1,3$ and $ {\bf e}_{\tau}:=
(0,1)^t$ for $\tau=2,4$ and
\begin{equation}
N_+(\bp) = \sqrt{2 E(\bp)(E(\bp)-1)}, \,\,\,\, N_-(\bp) = \sqrt{2 E(\bp)(E(\bp)+1)}.
\end{equation}
The indices $1$ and $2$ refer to positive ``eigenvalue'' $E(\bp)$ and
the indices $3$ and $4$ to negative $-E(\bp)$. (See, e.g., Evans et al.
\cite{Evansetal1996}.) Using Plancherel's theorem we get
\begin{multline}
  \label{v32}
  (\psi,P_3\psi)
  = \frac1{(2\pi)^{7}} \int_{\rz^3}
  d\bp_1\int_{\rz^3}d\bp_2\int_{\rz^3}d\bp_3\int_{\rz^3}d\bp_4
  \sum_{\tau_1,\tau_2,\tau_3,\tau_4 = 1}^4 \\
  \hat \chi(\bp_1 - \bp_2) \hat \varphi (\bp_2- \bp_3) \hat \varphi(\bp_3 -
  \bp_4)\hat
  \varphi(\bp_4-\bp_1)\times \\
  \times \la u_{\tau_1}( \bp_1)|u_{\tau_2}(\bp_2)\ra\la u_{\tau_2}(\bp_2)
  | u_{\tau_3}(\bp_3)\ra \la u_{\tau_3}(\bp_3) | u_{\tau_4}(\bp_4)\ra \la
  u_{\tau_4}(\bp_4) |
  u_{\tau_1}(\bp_1)\ra \\
  \times \dK \frac 1{(ia_{\tau_1} E(\bp_1) - \eta)(ia_{\tau_2} E(\bp_2)
    - \eta) (ia_{\tau_3} E(\bp_3) - \eta)(ia_{\tau_4} E(\bp_4) - \eta)},
\end{multline}
with $a_\tau = 1$ for $\tau=1,2$ and $a_\tau =- 1$ for $\tau=3,4$.
The integral over $\eta$ is seen to vanish by Cauchy's theorem, if all four
$a_{\tau_j}$ have the same sign. In fact we have to distinguish only two
cases, namely three of the $a_{\tau_j}$ are equal and two of the $a_{\tau_j}$
are equal.

Therefore in the following we will only treat two different cases. The
others then work analogously.

We begin with
\begin{equation}
  a_{\tau_1} = -1,\ a_{\tau_2} = a_{\tau_3} =a_{\tau_3} = 1.
\end{equation}
In that case the first factor in (\ref{v32}) reads
\begin{multline}
  \label{v33}
  \sum_{\tau_1=3,4}\la u_{\tau_1}(
  \bp_1)|u_{\tau_2}(\bp_2)\ra\sum_{\tau_2=1,2}\la u_{\tau_2}(\bp_2) |
  u_{\tau_3}(\bp_3)\ra\times \\ \times \sum_{\tau_3=1,2} \la
  u_{\tau_3}(\bp_3) | u_{\tau_4}(\bp_4)\ra\sum_{\tau_4=1,2}\la
  u_{\tau_4}(\bp_4) |
  u_{\tau_1}(\bp_1)\ra=\\
  \tr_{\cz^2} \Big[ \frac{\sigma\cdot \bp_1 \sigma\cdot \bp_2 +
    (1+E(\bp_1))(1-E(\bp_2))}{N_-(\bp_1)^2N_+(\bp_2)^2N_+(\bp_3)^2N_+(\bp_4)^2}
  \big[\sigma\cdot \bp_2 \sigma\cdot \bp_3 +
  (1-E(\bp_2))(1-E(\bp_3))\big]  \\
  \times \big[\sigma\cdot \bp_3 \sigma\cdot \bp_4 +
  (1-E(\bp_3))(1-E(\bp_4))\big] \big[\sigma\cdot \bp_4 \sigma\cdot \bp_1 +
  (1-E(\bp_4))(1+E(\bp_1))\big]\Big].
\end{multline}
We estimate the modulus of \eqref{v33} and obtain
\begin{eqnarray*}
  |(\ref{v33})| &\leq &\rc \frac{\tr_{\cz^2}|\sigma\cdot \bp_4
    \sigma\cdot
    \bp_1 + (1-E(\bp_4))(1+E(\bp_1))|}{N_-(\bp_1)N_+(\bp_4)} \\
  &\leq &\rc \frac{| \bp_4\cdot \bp_1 - (E(\bp_4)-1)(1+E(\bp_1))| +
    |\bp_4\wedge\bp_1|}{N_-(\bp_1)N_+(\bp_4)}.
\end{eqnarray*}
(Here and in the following $\rc$ is a generic positive constant.)
Since
\begin{multline*}
  \frac1{2\pi}\dK \frac 1{(-i E(\bp_1) - \eta)(i E(\bp_2) - \eta)
    (i E(\bp_3) - \eta)(i E(\bp_4) - \eta)} \\
  = \frac 1{(E(\bp_2) + E(\bp_3))(E(\bp_2) + E(\bp_3))(E(\bp_3) +
    E(\bp_4))}
\end{multline*}
our term of interest (\ref{v32}) is bounded by constant times
\begin{multline}\label{v34}
  \int_{\rz^3} d\bp_1\int_{\rz^3}d\bp_2\int_{\rz^3}d\bp_3\int_{\rz^3}d\bp_4
  |\hat \chi(\bp_1 - \bp_2) \hat \varphi (\bp_2- \bp_3) \hat \varphi(\bp_3
  - \bp_4) \hat
  \varphi(\bp_4- \bp_1)|\\
  \times \frac{| \bp_4\cdot \bp_1 - (E(\bp_4)-1)(E(\bp_1)+1)| + |\bp_4
    \wedge \bp_1|}{N_-(\bp_1)N_+(\bp_4)(E(\bp_2) + E(\bp_3))(E(\bp_2) +
    E(\bp_3))(E(\bp_3) + E(\bp_4) }).
\end{multline}
Substituting $\bp_2 \to \bp_1 + \bp_2$ turns (\ref{v34}) into
\begin{multline}\label{v35}
  \int_{\rz^3}
 d\bp_1\int_{\rz^3}d\bp_2\int_{\rz^3}d\bp_3\int_{\rz^3}d\bp_4\\
 |\hat \chi( - \bp_2)|\frac{| \hat
    \varphi (\bp_2 + \bp_1- \bp_3) \hat \varphi(\bp_3 - \bp_4) \hat
    \varphi(\bp_4-\bp_1)|}{N_-(\bp_1)N_+(\bp_4)} \\
  \times \frac {| \bp_4\cdot \bp_1 - (E(\bp_4)-1)(E(\bp_1)+1)| + |\bp_4
    \wedge \bp_1|}{(E(\bp_2+\bp_1) + E(\bp_3))(E(\bp_2+\bp_1) +
    E(\bp_3))(E(\bp_3) + E(\bp_4) }\\
  = \int_{\rz^3} d\bp_2 |\hat \chi(-\bp_2)| f(\bp_2)
\end{multline}
where we introduce $f$ to be the remaining integrand. 
We will now estimate $f$. Substituting $\bp_1 \to \bp_1 +
\bp_4$, $\bp_3 \to \bp_3 + \bp_4$ we get
\begin{multline*}
  f(\bp_2) = \int_{\rz^3} d\bp_1\int_{\rz^3}d\bp_3\int_{\rz^3}d\bp_4
\hat \varphi(\bp_2+\bp_1-\bp_3) \hat \varphi(\bp_3) \hat \varphi(\bp_1)| \\
  \times \frac {| \bp_4\cdot (\bp_1 +\bp_4) -
    (E(\bp_4)-1)(1+E(\bp_1+\bp_4))| + |\bp_4 \wedge
    \bp_1|}{N_-(\bp_1+\bp_4)N_+(\bp_4)(E(\bp_2+\bp_1+\bp_4) + E(\bp_3+\bp_4)) }\\
  \times \frac 1{(E(\bp_2+\bp_1+\bp_4) + E(\bp_3+\bp_4))(E(\bp_3+\bp_4) +
    E(\bp_4))}.
\end{multline*}
Since
$$
  E(\bp_1 + \bp_4) = E(\bp_4) + \frac{\bp_4 \cdot\bp_1}{E(\bp_4)}\mu 
$$
for some $\mu \in [0,1]$, 
we see that
$$
  | \bp_4\cdot (\bp_1 +\bp_4) - (E(\bp_4)-1)(E(\bp_1+\bp_4)+1)| + |\bp_4
  \wedge \bp_1|\leq 4 |\bp_1||\bp_4|.
$$
Notice, we can bound
$$
\int_{\rz^3} d\bp_4 \frac{|\bp_4|}{E(\bp_3+ \bp_4) E(\bp_4)^2 N_+(\bp_1 +
  \bp_4)} 
\leq \rc
$$
independent of $\bp_1$ and $\bp_3$.
Therefore,
$$
f(\bp_2) \leq \rc \int_{\rz^3} d\bp_1\int_{\rz^3} d\bp_3 |\hat \varphi (\bp_2 +
\bp_1- \bp_3) \hat \varphi(\bp_3) \hat \varphi(\bp_1)| |\bp_1|.
$$
Since $\hat \varphi(\bk) =4\pi \hat n(\bk)/\bk^2$, we have that $f(0)$
is finite; since $\hat n$ is compactly supported, $\bp_1$ and $\bp_2$
are bounded. We conclude that $f$ has also compact support.

Consequently, since
\begin{equation}
 \int_{\rz^3} d\bp_2 |\hat\chi(-\bp_2)|f(\bp_2) \leq \|\hat\chi\|_q \|f\|_p,
\end{equation}
we see using the Hausdorff-Young inequality that for all $p\geq 2$
\begin{equation}\label{v37}
 \int_{\rz^3} d\bp_2 |\chi(-\bp_2)|f(\bp_2) \leq \rc_{p,\varphi} \|\chi\|_p
\end{equation}
for constant $\rc_{p,\varphi}$ depending on $p$ and $\varphi$.

Next, we take a peek at the case $ a_{\tau_1} = a_{\tau_2}=1$ and
$a_{\tau_3} =a_{\tau_4} = -1$.  The corresponding integral over $\eta$
gives
\begin{multline*}
  \frac1{2\pi}\dK \frac 1{(i E(\bp_1) - \eta)(i E(\bp_2) - \eta)
    (-i E(\bp_3) - \eta)(-i E(\bp_4) - \eta)} \\
  = \frac 1{(E(\bp_2) + E(\bp_3))(E(\bp_2) +
    E(\bp_4))(E(\bp_1) + E(\bp_4))} \\
  + \frac 1{(E(\bp_2) + E(\bp_3))(E(\bp_1) + E(\bp_3))(E(\bp_1) +
    E(\bp_4))}.
\end{multline*}
Observe now that the corresponding first factor in (\ref{v32}) can be
bounded by $\rc \cdot 4 \frac{|\bp_2||\bp_3|}{N_+(\bp_2)N_-(\bp_3)}$. Now,
we do similar variable transforms as above and arrive
at an analogue of (\ref{v37}).
\end{proof}
Again this has the
\begin{corollary}
  \label{sec:proof-self-adjo-1}
  The perturbation $P_3$ is relatively form bounded with respect to $|D^0|$
  with form bound zero.
\end{corollary}
The above result was the final step in showing that all four
perturbations, $U$, $P_3$, $P_4$, and $X$, are form bounded with
respect to $\Pvp\dv\Pvp$.


\end{document}